\newcommand{\Tr}{\mathop{\mathrm{Tr}}\nolimits}
\newcommand{\matriz}[1]{{\bm{\mathsf{#1}}}}  

\documentclass[prl,twocolumn,superscriptaddress,showpacs]{revtex4}

\usepackage{amsfonts,amssymb,amsbsy,bm,graphicx,times}

\begin{document}

\title{Escherlike quasiperiodic heterostructures}

\author{Alberto~G.~Barriuso}
\affiliation{Departamento de \'Optica, 
Facultad de F\'{\i}sica, Universidad Complutense, 
28040~Madrid, Spain}

\author{Juan~J.~ Monz\'on}

\affiliation{Departamento de \'Optica, 
Facultad de F\'{\i}sica, Universidad Complutense, 
28040~Madrid, Spain}

\author{Luis~L.~S\'{a}nchez-Soto}
\affiliation{Departamento de \'Optica, 
Facultad de F\'{\i}sica, Universidad Complutense, 
28040~Madrid, Spain}

\author{Antonio~F.~Costa} 
\affiliation{Departamento de Matem\'{a}ticas Fundamentales, 
Facultad de Ciencias, 
Universidad Nacional de Educaci\'{o}n a Distancia, 
Senda del Rey 9, 28040 Madrid, Spain}

\bigskip

\date{\today}

\begin{abstract}
We propose quasiperiodic heterostructures associated with the
tessellations of the unit disk by regular hyperbolic triangles.
We present explicit construction rules and explore some of the 
properties exhibited by these geometric-based systems. 

\end{abstract}

\maketitle

Quasiperiodic (QP) systems have been receiving a lot of attention 
over the last years~\cite{Macia:2006}. The interest was originally
motivated by the theoretical predictions that they should manifest
peculiar electron and phonon critical states~\cite{Ostlund:1984,
Kohmoto:1987}, associated with highly fragmented fractal energy
spectra~\cite{Kohmoto:1983,Suto:1989,Bellissard:1989}.  On the other
hand, the practical fabrication of Fibonacci~\cite{Merlin:1985} and
Thue-Morse~\cite{Merlin:1987} superlattices has triggered a number of
experimental achievements that have provided new insights into the
capabilities of QP structures~\cite{Velasco:2003}.  In particular,
possible optical applications have deserved major attention and some
intriguing properties have been
demonstrated~\cite{Tamura:1989,Vasconcelos:1999,Lusk:2001,Barriuso:2005}.
Underlying all these theoretical and experimental efforts a crucial
fundamental question remains concerning whether QP devices would
achieve better performance than usual periodic ones for some specific
applications~\cite{Macia:2001}.

The QP systems considered thus far rely for their explicit
construction in substitutional rules among the elements of a basic
alphabet.  In the common case of a two-letter alphabet $\{A, B \}$,
the algorithm takes the form $A \mapsto \sigma_{A} (A, B)$, $B
\mapsto \sigma_{B} (A, B)$, where $\sigma_{A}$ and $\sigma_{B}$
can be any string of the letters. The sequences generated after $n$
applications of the algorithm are of significance in fields as diverse
as cryptography, time-series analysis, and cellular
automata~\cite{Cheng:1990}.  In addition, they have interesting
algebraic properties, which are usually characterized by the nature of
their Fourier or multifractal spectra~\cite{Spinadel:1999}.

We wish to approach the problem from an alternative geometrical
perspective. To this end, we first observe that in many problems 
of physical interest~\cite{Macia:2006b} the letters of the alphabet 
can be identified with one-dimensional linear lossless systems 
(i.e., with two input and two output channels).  Under these general 
conditions, it turns out that the associated transfer matrix belongs 
to the group SU(1,1), which is also the basic symmetry group of the 
hyperbolic geometry~\cite{Coxeter:1968}. In consequence, the unit 
disk appears as the natural arena to discuss their
performance~\cite{Yonte:2002,Monzon:2002,Barriuso:2003}.  Since in the
Euclidean plane, QP behavior is intimately linked with tessellations,
one is unfailingly led to consider the role of hyperbolic
tessellations in the unit disk, much in the spirit of Escher's
masterpiece woodcut \textit{Circle Limit~III}~\cite{Coxeter:1996}.
The answer we propose is promising: the tessellations by different
regular polygons provide new sequences with properties that may open
avenues of research in this field.

A QP system can thus be seen as a word generated by stacking
different letters of the basic alphabet. To be specific, we focus our
attention on the optical response. Let us consider one of these
letters (which in practice is made of several plane-parallel layers),
which we assume to be sandwiched between two semi-infinite identical
ambient ($a$) and substrate ($s$). We suppose monochromatic plane
waves incident, in general, from both the ambient and the
substrate. As a result of multiple reflections in all the interfaces,
the total electric field can be decomposed in terms of forward- and
backward-traveling plane waves, denoted by $E^{(+)}$ and $E^{(-)}$,
respectively. If we take these components as a vector
\begin{equation}
  \label{Evec}
  \mathbf{E} = 
  \left ( 
    \begin{array}{c}
      E^{(+)} \\ 
      E^{(-)} 
    \end{array}
  \right ) \, ,
\end{equation}
then the amplitudes at both the ambient and the substrate sides  
are related by the transfer matrix $\matriz{M}$
\begin{equation}
  \label{M1}
  \mathbf{E}_a =  
  \matriz{M} \,
  \mathbf{E}_s \, .
\end{equation}
It can be shown that $\matriz{M}$ is of the form
\begin{equation}
  \label{Mlossless}
  \matriz{M} =
  \left (
    \begin{array}{cc}
      1/T & R^\ast/T^\ast \\ 
      R/T & 1/T^\ast
    \end{array}
  \right )  \, ,
\end{equation}
where the complex numbers $R$ and $T$ are, respectively, the overall
reflection and transmission coefficients for a wave incident from the
ambient.  The condition $\det \matriz{M} = +1$ is equivalent to $|R|^2
+ |T|^2 = 1$, and then the set of transfer matrices reduces to the
group SU(1,1). Obviously, the matrix of a word obtained by putting
together letters of the alphabet is the product of the matrices
representing each one of them, taken in the appropriate order.

In many instances we are interested in the transformation properties
of field quotients rather than the fields themselves. Therefore, it
seems natural to consider the complex numbers
\begin{equation}
  z = \frac{E^{(-)}}{E^{(+)}} \, ,
\end{equation}
for both ambient and substrate.  The action of the matrix given in
Eq.~(\ref{M1}) can be then seen as a function $z_a = f(z_s)$ that
can be appropriately called the transfer function. From a geometrical
viewpoint, this function defines a transformation of the complex plane
$\mathbb{C}$, mapping the point $z_s$ into the point $z_a$ according
to
\begin{equation}
  \label{accion}
  z_a =  
  \frac {\beta^\ast + \alpha^\ast z_s} {\alpha + \beta z_s} \, ,
\end{equation}
where $\alpha = 1/T$ and $\beta = R ^\ast/T^\ast$. When no light is
incident from the substrate, $z_s = 0$ and then $z_a = R$.  Equation
(\ref{accion}) is a bilinear (or M\"{o}bius) transformation.  One can
check that the unit disk, the external region and the unit circle
remain invariant under (\ref{accion}). This unit disk is then a model
for hyperbolic geometry in which a line is represented as an arc of a
circle that meets the boundary of the disk at right angles to it (and
diameters are also permitted). In this model, we have three different
kinds of lines: intersecting, parallel (they intersect at infinity,
which is precisely the boundary of the disk) and ultraparallel (they
are neither intersecting nor parallel).

To classify the possible actions it proves convenient to work out
the fixed points of the transfer function; that is, the field
configurations such that $z_a = z_s \equiv z_f$ in Eq.~(\ref{accion}),
whose solutions are
\begin{equation}
  z_f = \frac{1}{2 \beta} \left \{  -2 i \, \mathrm{Im}(\alpha) \pm
    \sqrt{[ \Tr ( \matriz{M} )]^2 -4} \right \} .
\end{equation}
When $ [\Tr ( \matriz{M} )] ^2 < 4$ the action is elliptic and it has
only one fixed point inside the unit disk. Since in the Euclidean
geometry a rotation is characterized for having only one invariant
point, this action can be appropriately called a hyperbolic rotation.

When $ [ \Tr ( \matriz{M} )]^2 > 4$ the action is hyperbolic and it
has two fixed points, both on the boundary of the unit disk. The
geodesic line joining these two fixed points remains invariant and
thus, by analogy with the Euclidean case, this action is called a
hyperbolic translation.

Finally, when $ [ \Tr (\matriz{M}) ]^2 = 4$ the action is parabolic
and it has only one (double) fixed point on the boundary of the unit
disk.

As it is well known, SU(1, 1) is isomorphic to the group of real
unimodular matrices SL(2, $\mathbb{R}$), which allows us to translate
the geometrical structure defined in the unit disk to the complex
upper semiplane, recovering in this way an alternative model of the
hyperbolic geometry that is useful in some applications.

The notion of periodicity is intimately connected with tessellations,
i.e., tilings by identical replicas of a unit cell (or fundamental
domain) that fill the plane with no overlaps and no gaps.  Of special
interest is the case when the primitive cell is a regular polygon with
a finite area~\cite{Zieschang:1980}. In the Euclidean plane, the
associated regular tessellation is generically noted $\{p, q \}$,
where $p$ is the number of polygon edges and $q$ is the number of
polygons that meet at a vertex. We recall that geometrical constraints
limit the possible regular tilings $\{p, q \}$ to those verifying $(p
- 2)(q - 2) = 4$. This includes the classical tilings $ \{4, 4\}$
(tiling by squares) and $\{ 6, 3 \}$ (tiling by hexagons), plus a
third one, the tiling $\{3, 6\}$ by triangles (which is dual to the
$\{6, 3\}$).

On the contrary, in the hyperbolic disk regular tilings exist
provided $(p - 2)(q - 2) > 4$, which now leads to an infinite number
of possibilities. An essential ingredient is the way to obtain
fundamental polygons. These polygons are directly connected to the
discrete subgroups of isometries (or congruent mappings). Such groups
are called Fuchsian groups~\cite{Ford:1972} and play for the
hyperbolic geometry a role similar to that of crystallographic groups
for the Euclidean geometry~\cite{Beardon:1983}.

A tessellation of the hyperbolic plane by regular polygons has a
symmetry group that is generated by reflections in geodesics, which
are inversions across circles in the unit disk.  These geodesics
correspond to edges or axes of symmetry of the polygons. Therefore, to
construct a tessellation of the unit disk one just has to built one
tile and to duplicate it by using reflections in the edges.

\begin{figure}
  \centering
  \resizebox{0.90\columnwidth}{!}{\includegraphics{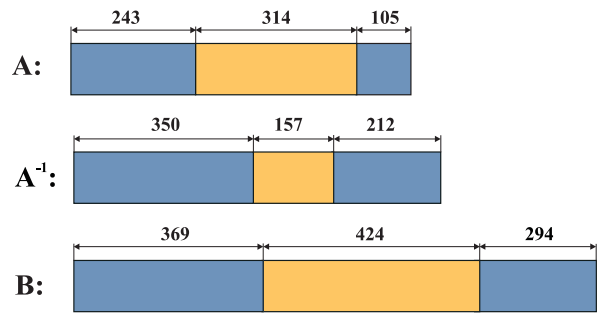}}
  \caption{(color online). A realistic implementation of the
    generators $\matriz{A}$ and $\matriz{B}$ (together with
    $\matriz{A}^{-1}$) of the tessellations by hyperbolic triangles in
    (\ref{mattrian}). The external layers (in blue) are made of
    cryolite (Na$_3$AlF$_6$), while the central medium (in brown) is
    zinc selenide (ZnSe). The wavelength in vacuum is $\lambda =
    610$~nm and normal incidence (from left to right) has been
    assumed.  The corresponding thicknesses are expressed in
    nanometers.}
\end{figure}

In this Letter, we consider only the simple example of a tessellation by
triangles with vertices in the unit circle, although the treatment can
be extended to other polygons.  The key idea is to consider the
Fuchsian group generated by an elliptic transformation whose fixed
point is the middle of an edge of the triangle and a parabolic one
with its fixed point in the opposite vertex.  Proceeding in this way
we get 
\begin{eqnarray}
   \label{mattrian}
 \matriz{A} &  = & 
  \left (
    \begin{array}{cc}
      1 + i/\sqrt{3} &  1/\sqrt{3} \\
      1/\sqrt{3} & 1 - i/\sqrt{3}
    \end{array}
  \right ) \, , 
  \nonumber \\
  \\
  \matriz{B} & =  & \left (
    \begin{array}{cc}
      2i/\sqrt{3} &  -1/\sqrt{3} \\
     
      -1/\sqrt{3} & -2i/\sqrt{3}
    \end{array}
  \right ) \, . \nonumber
\end{eqnarray}
The fixed point of $\matriz{A}$ is $-i$; while for $\matriz{B}$ the
fixed point in the disk is $ i (2 - \sqrt{3})$.  In Fig.~1 we show a
possible way in which these matrices can be implemented in terms
of two commonly employed materials in optics.  Note that, in physical
terms, the inverses must be constructed as independent systems,
although in our case only $\matriz{A}^{-1}$ must be considered, since
the action of $\matriz{B}^{-1}$ coincides with that of $\matriz{B}$.

\begin{figure}
  \centering
  \resizebox{0.85\columnwidth}{!}{\includegraphics{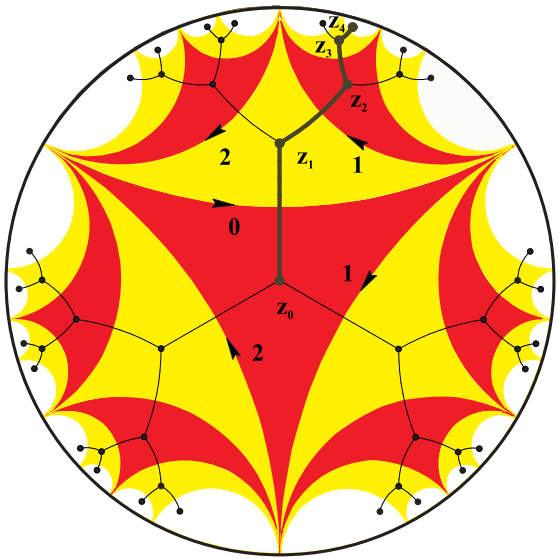}}
  \caption{(color online). Tiling of the unit disk with the matrices
    (\ref{mattrian}).  The marked points are the barycenters of the
    triangles in the tessellation and all of them are the transformed
    of the origin by a matrix that have as reflection coefficient the
    complex number that links the origin with the center of the
    triangle.}
\end{figure}

\begin{table}[b]
  \caption{Explicit rules to obtain the barycenter $z_{n+1}$ from
    the $z_{n}$. We have indicated the corresponding transformations,
    which depend on the color jumps and the sides crossed by going from
    $z_{n}$ to  $z_{n+1}$. \label{unica}}
  \begin{ruledtabular}
    \begin{tabular}{llll|lll}
      &\multicolumn{3}{c}{red $\rightarrow $ yellow} & 
      \multicolumn{3}{c}{yellow $\rightarrow $ red}\\
      Side & Trans & $ \matriz{A}_{n+1}$ & $\matriz{B}_{n+1}$ & 
      Trans & $\matriz{A}_{n+1}$ & $\matriz{B}_{n+1}$ \\ \hline
      0 & $\matriz{B}_{n}$ & $\matriz{B}_{n} \matriz{A}_{n} \matriz{B}_{n}$ & 
      $\matriz{B}_{n}$ & $\matriz{B}_{n}$ & 
      $\matriz{B}_{n} \matriz{A}_{n}^{-1} \matriz{B}_{n}$ & $\matriz{B}_{n} $  \\
      1 & $\matriz{A}_{n}$ & $\matriz{A}_{n}$ & 
      $\matriz{A}_{n} \matriz{B}_{n} \matriz{A}_{n}^{-1}$ & $\matriz{A}_{n}^{-1}$ & 
      $\matriz{A}_{n}$ & $\matriz{A}_{n}^{-1} \matriz{B}_{n} \matriz{A}_{n}$   \\
      2 &  $\matriz{A}_{n}^{-1}$ & $ \matriz{A}_{n}^{-1}$ & 
      $\matriz{A}_{n}^{-1} \matriz{B}_{n} \matriz{A}_{n}$ &  $\matriz{A}_{n}$ & 
      $ \matriz{A}_{n}^{-1}$ & $ \matriz{A}_{n} \matriz{B}_{n} \matriz{A}_{n}^{-1}$ 
    \end{tabular}
  \end{ruledtabular}
\end{table} 

In Fig. 2 we have shown the tessellation obtained by transforming the
fundamental triangle with the Fuchsian group generated by the powers
of $\{ \matriz{A}, \matriz{B} \}$ (and the inverses). This triangle is
equilateral with vertices at the points $- i$, $\exp (i \pi /6)$ and
$\exp (i 5 \pi /6)$ (which are the fixed points of $\matriz{A}$,
$\matriz{A} \matriz{B}$ and $\matriz{B} \matriz{A}$, respectively).
Moreover, all the other triangles are equal, with an area $\pi$. In
the figure we have plotted also the barycenters of each triangle
together with the resulting tree (that is called the dual graph of the
tessellation), which turns out to be a Farey
tree~\cite{Schroder:2006}. In fact, each line connecting two of these
barycenters represent the action of a word (with alphabet $\{
\matriz{A}, \matriz{B} \}$ and the inverses).

To give an explicit construction rule for the possible words, we
proceed as follows. First, we arbitrarily assign the number 0 to the
upper side of the fundamental triangle, while the other two sides are
clockwise numbered as 1 and 2. It is easy to convince oneself that
this assignment fixes once for all the numbering for the sides of the
other triangles in the tessellation. However, these triangles can be
distinguished by their orientation (as seen from the corresponding
barycenter): the clockwise oriented are filled in red, while the
counterclockwise are filled in yellow. In short, we have determined a
fundamental coloring of the tessellation~\cite{Grunbaum:1987}.

To obtain one barycenter $z_{n+1}$ from the previous one $z_{n}$, one
looks first at the corresponding color jump. Next, the matrix that
take $z_{n}$ into $z_{n+1}$ depends on the numbering of the side (0,
1, or 2) one must cross, and appear in the appropriate column
``Trans'' in Table~\ref{unica}. The next generation is obtained much
in the same way, except for the fact that $\matriz{A}_{n}$ and
$\matriz{B}_{n}$ must be replaced by $\matriz{A}_{n+1}$ and
$\matriz{B}_{n+1}$, respectively, as indicated in the Table. In
obtaining recursively any word, the origin is denoted as $z_0$ and 
the matrices $\matriz{A}_{0}$ and $\matriz{B}_{0}$ coincide with 
$\matriz{A}$ and $\matriz{B}$.

With this rule, one can construct any word proceeding step by
step. For example, the word that transform $z_0$ into $z_5$ in 
the zig-zag path sketched in Fig.~2
results
\begin{equation}
  \label{eq:1}
  \begin{array}{lcl}
    z_{0} \rightarrow z_{1} \quad : \quad \matriz{B} \, ,\\
    z_{1} \rightarrow z_{2} \quad : \quad 
    \matriz{B} \matriz{A}^{-1} \matriz{B}  \, , \\
    z_{2} \rightarrow z_{3} \quad : \quad 
    \matriz{B}  \matriz{A}^{-1} \matriz{B}  \, ,\\
    z_{3} \rightarrow z_{4} \quad : \quad 
    \matriz{B}  \matriz{A}^{-1}  \matriz{A}^{-1} 
    \matriz{B}  \matriz{A} \matriz{A} \matriz{B}  \, , \\ 
    z_{4} \rightarrow z_{5} \quad : \quad 
    \matriz{B}  \matriz{A}^{-1}  \matriz{A}^{-1} 
    \matriz{B}  \matriz{A}  \matriz{B} \matriz{A} \matriz{A} \matriz{B} \, .
  \end{array}
\end{equation}
Obviously, the total word is obtained by composing these partial
words.

In fact, one can show that given an arbitrary sequence of nonzero
integers $\{k_{1}, k_{2}, \ldots, k_{r} \}$, the word represented by
the transfer matrix
\begin{equation}
  \label{eq:Antoine}
  \matriz{M} (s, k_{1}, \ldots, k_{r}) = \matriz{B}^{s_1}  
  \matriz{A}^{k_{1}}  \matriz{B} \matriz{A}^{k_{2}}
  \ldots  \matriz{B} \matriz{A}^{k_{r}} \matriz{B}^{s_2} \, ,
\end{equation}
where $\{s_{1},s_{2}\} \subset \{0,1\}$, transforms the origin in a
barycenter of the tessellation.

\begin{figure}
  \centering
  \resizebox{0.95\columnwidth}{!}{\includegraphics{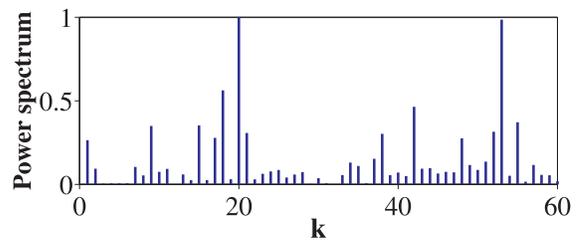}}
  \caption{(color online). Normalized structure factor for the word
    (made of 60 letters) connecting the origin with the point $z_8$ in
    the zig-zag path shown in Fig. 2.}
\end{figure}

Given the geometric regularity of the construction sketched in
this Letter, the sequences obtained must play a key role in the theory
and practice of QP systems. Of course, to put forward the relevant
physical features of these sequences, there are a number of quantities
one can look at. Perhaps one of the most appropriate ones to assess
the performance of these systems is the structure factor~\cite{Cheng:1990}.  
For a word $\matriz{M}$ (composed of $L$ letters) we define a numerical 
sequence $f_n$ by assigning  1, $ e^{2 \pi i/3 }$, and $ e^{-2 \pi i/3 }$ to 
the letters $\matriz{B}$,  $\matriz{A}$, and $\matriz{A}^{-1}$, respectively. 
Next, we calculate the discrete Fourier transform of the sequence $f_n$
\begin{equation}
  F_k = \sum_{n=0}^{L-1} f_{n} \exp \left ( - \frac{2 \pi i k n}{L} \right ) \, ,
\end{equation}
where $k=0,1,..L-1$. The structure factor (or power spectrum) is just
$|F_k|^2$.  In Fig.~3 we have plotted this structure factor in terms
of $k$ for the word connecting the origin with the point $z_{8}$ in
the zig-zag path of Fig.~2. The peaks reveal a rich behavior: a full
analysis of these questions is outside the scope of this Letter and
will be presented elsewhere.

As a final and rather technical remark, we note that the quotient of
the hyperbolic disk by the Fuchsian group generated by $\matriz{A}$
and $\matriz{B}$ is a 2-orbifold of genus 0, with a conical point of
order two and a cusp.  Each word as given in Eq.~(\ref{eq:Antoine})
represents a hyperbolic transformation of the disk, and the axis of
the transformation is projected onto a closed geodesic of such an
orbifold. This provides an orbifold interpretation of our QP
sequences.

In summary, we expect to have presented new schemes to generate QP
sequences based on hyperbolic tessellations of the unit disk. Apart
from the intrinsic beauty of the formalism, our preliminary results
seem to be quite encouraging for future applications of these systems.

The authors wish to express their warmest gratitude to E. Maci\'a and
J. M. Montesinos for their help and interest in the present work.


\end{document}